\documentclass{sf2a-conf}
\usepackage{graphicx}
\usepackage{natbib}
\begin{document}

\TitreGlobal{SF2A 2008}

\title{Rising optical afterglows seen by TAROT}
\author{Gendre, B.}\address{Laboratoire d'Astrophysique de Marseille/CNRS/Université de Provence}
\author{Klotz, A.}\address{Centre d'Etude Spatiale des Rayonnements/CNRS/Université Paul Sabatier}
\author{Stratta, G.}\address{ASI Science Data Center}
\author{Preger, B.$^3$}
\author{Piro, L.}\address{IASF-Roma}
\author{Pelangeon, A.}\address{Laboratoire d'Astrophysisque de Toulouse/CNRS}
\author{Galli, A.$^4$}
\author{Cutini, S.$^3$}
\author{Corsi, A.$^4$}
\author{Bo\"er, M.}\address{Observatoire de Haute Provence/CNRS}
\author{Atteia, J.L.$^5$}
\runningtitle{Rising optical afterglows seen by TAROT}
\setcounter{page}{237}

\index{Gendre, B.}
\index{Stratta, G.}
\index{Preger, B.}
\index{Piro, L.}
\index{Pelangeon, A.}
\index{Galli, A.}
\index{Cutini, S.}
\index{Corsi, A.}
\index{Bo\"er, M.}
\index{Atteia, J.L.}

\maketitle
\begin{abstract} We present the multi-wavelength study of those gamma-ray bursts observed by TAROT. These events are characterized by the presence at early time of a rising in their optical light curves lasting a few hundred of seconds. In one case (GRB 060904B), a flare occurs at similar time in the X-ray band, while in the other cases the X-ray light curves appear smooth during the optical rise. We investigate the possible nature of this behavior and conclude that a multi-component emission is mandatory to explain the optical-to-X-ray afterglow.  \end{abstract}
%
\section{Introduction}

Since the discovery of Gamma-Ray Burst (GRB) afterglows, in 1997 \citep{costa97}, tens of GRB optical afterglows have been detected by ground-based rapid response telescopes. Early optical afterglow data play a relevant role to obtain information on the physics of the central engine, and possibly to constraint the initial Lorentz factor of the fireball \citep[{\it e.g} ][]{zhang03}. This paper is devoted to the analysis of the GRB observations made by TAROT during the period 2001-2007 when the first telescope started to be fully operational.

\section{GRB observations with TAROT}
\label{Observations}

\begin{figure*}[htb]
\begin{center}
\begin{tabular}{c}
\includegraphics[width=6cm]{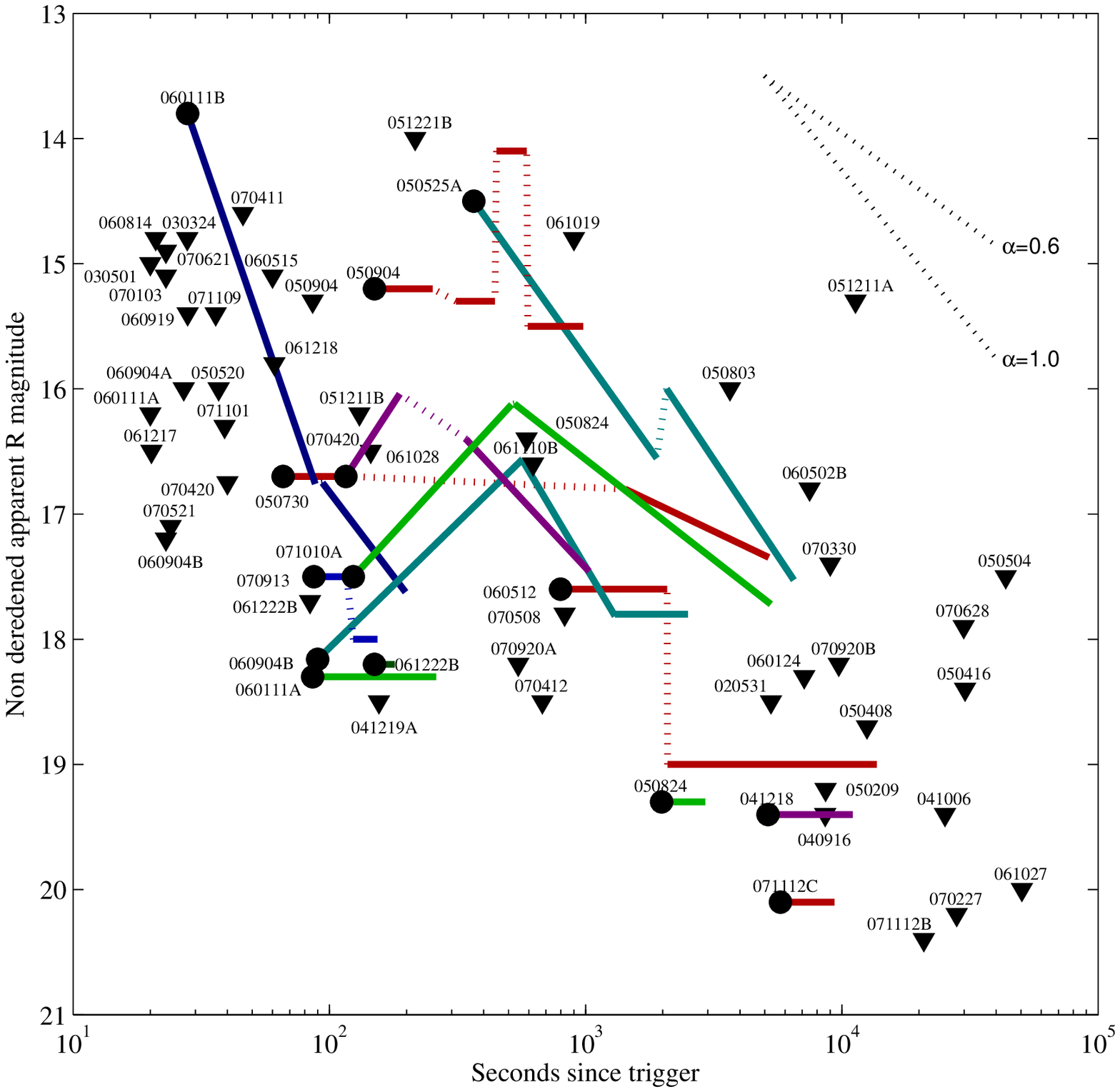}
\includegraphics[width=5cm]{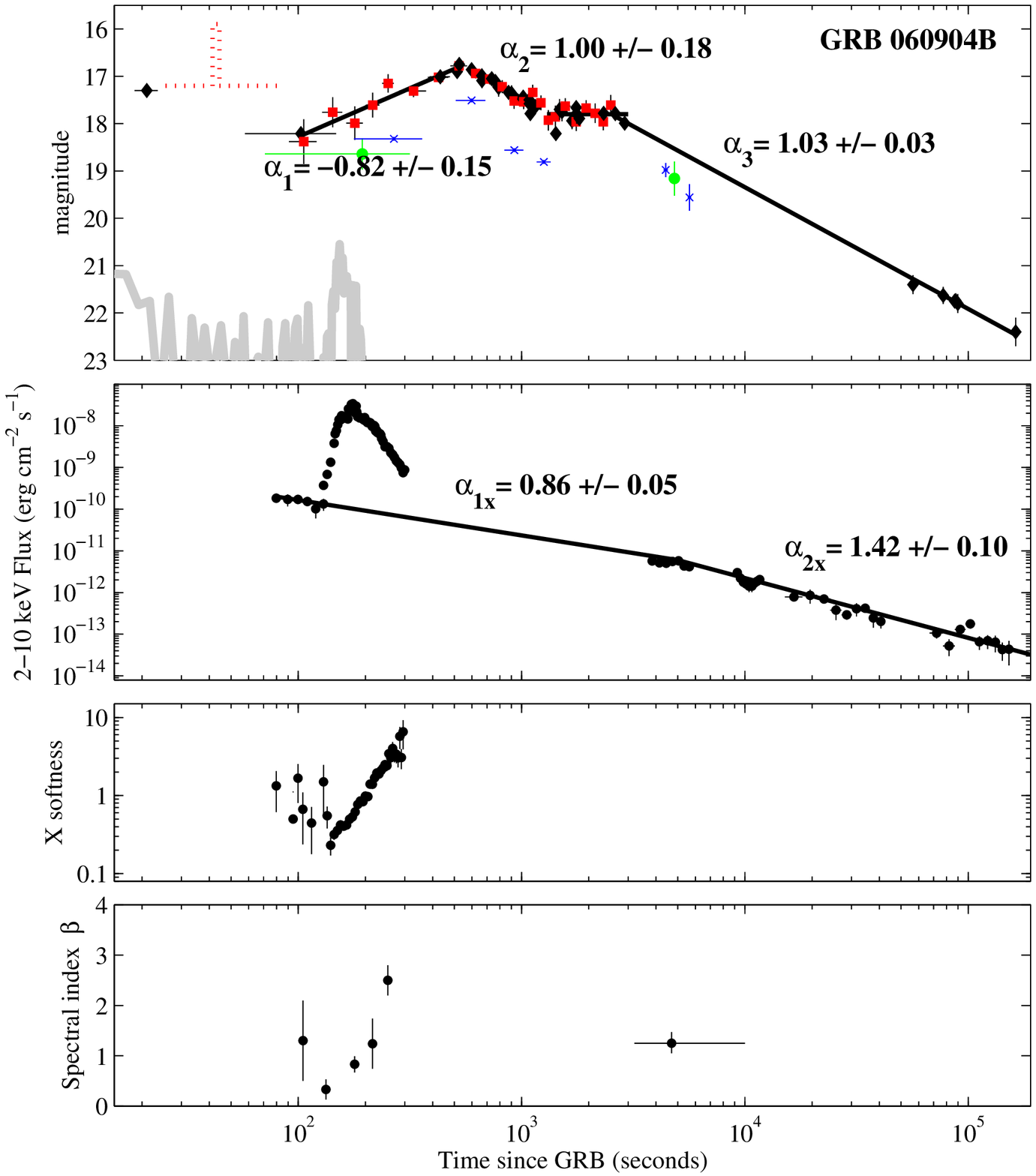}
\includegraphics[width=6cm]{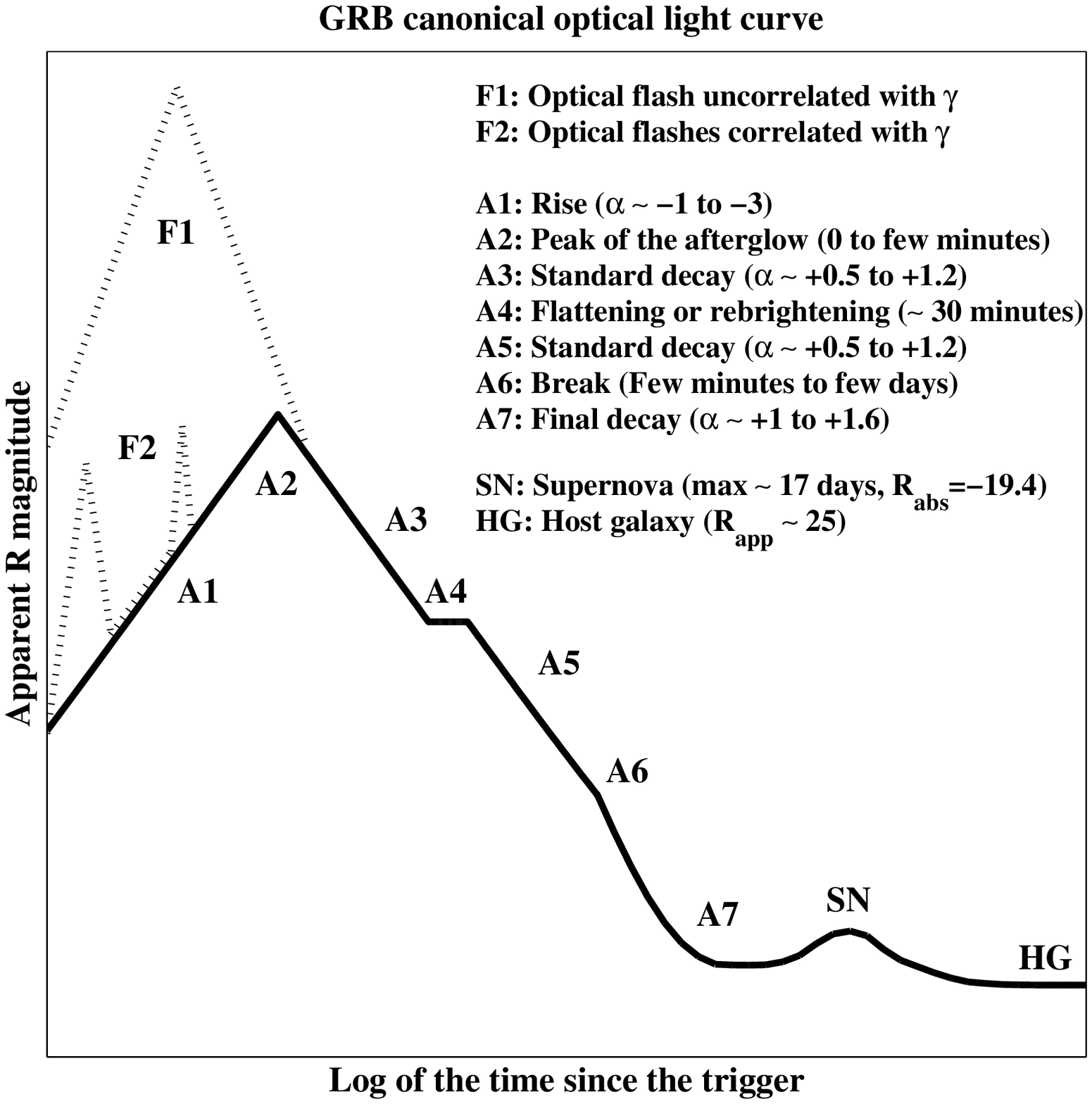}
\end{tabular}
\caption{
Left : the TAROT data: upper limits for the first time bin (triangles)
and light curves  of the detected sources (thick lines).
The upper right slopes illustrate some typical afterglow decays.
Middle : Time resolved parameters of GRB~060904B.
Top panel: Optical light-curve. 
Red squares are TAROT data (plus the limiting
magnitude during 23s to 83s indicated by the
dotted line).
Black diamonds are R measurements of other
observatories. Green disk and blue x are
from literature for V and B band respectively.
The BAT light curve is displayed as the light gray curve
offseted arbitrarily.
Fits are based on a power law.
Second panel: X-Ray light-curve from XRT data.
The count rate has been converted to flux units
using the best fit spectral model of late afterglow.
Third panel: softness ratio defined by (0.3-1.5 keV)/(1.5-10 keV).
Bottom panel: X-ray band spectral index.
Right : Canonical optical light curve of a typical GRB.
}
\label{tarot_firsts}
\end{center}
\end{figure*}

TAROT observed 59~GRBs between 2001 and 2007. During that period, 13 optical transients were detected. Six of them are time resolved allowing the measure of the decay index. 12 GRBs were observed when the gamma emission was still active amongst them 3 were positively detected by TAROT. Figure \ref{tarot_firsts} displays all the detected optical emissions and the first upper limit in the case of negative detections. Half of the afterglows with detectable OT exhibit an increase in brightness until few hundreds of seconds (GRB~050904, GRB~060904B, GRB~070420, GRB~071010A).

\section{A rising in detail : GRB 060904B}

The optical light-curve of GRB~060904B is shown in Fig.~\ref{tarot_firsts}. Before the end of the prompt phase,
the optical emission features a rise, reaching R=16.8 at maximum. During that rising part, the X-ray and gamma-ray emissions features a giant flare, well fit by a Band model and an hard-to-soft behavior, characteristic of prompt related emission \citep{klo08}. Several phenomena can contribute to a late rising of the optical afterglow \citep[see][]{klo08}, all of them imply a double component to explain the observed emission.

\section{Toward a canonical light curve}
\label{Canonical light curve}

Ten years of GRB optical observations allow to derive a global view of their light curves (presented in figure~\ref{tarot_firsts}). Components F1 and F2, two type of flash emissions, have been observed only when the GRB is still in its active phase: F1 variations are not correlated with the gamma--ray activity, while F2 emission follows the gamma--ray activity. Often, neither F1 nor F2 are detected. A1 is only observed when the peak of the afterglow A2 occurs tens to hundreds seconds seconds after the GRB. A3 is the usual afterglow power law decay. The A4 flattening is more rare and occurs usually between 15 and 30\,min in the source rest frame \citep[see discussion in ][]{Klotz2005}. The nature of the A6 break, previously supposed to be the jet break, is now under debate \citep[e.g.][]{Covino2006}. The SN component is the supernova light curve signature that has been detected only for the nearest sources. The final HG segment is the host galaxy background flux. TAROT observations are usually sensitivity limited to the A5 phase.

\begin{acknowledgements}
B. Gendre. acknowledges support from the CNES.
The TAROT telescope has been funded by the {\it Centre National
de la Recherche Scientifique} (CNRS), {\it Institut National des
Sciences de l'Univers} (INSU) and the Carlsberg Fundation. It has
been built with the support of the {\it Division Technique} of
INSU. We thank the technical staff contributing to the TAROT project:
G. Buchholtz, J. Eysseric, M. Merzougui, C. Pollas, and P. \& Y. Richaud.
S. Cutini, B. Preger \& G. Stratta acknowledge support
of ASI contract 1/024/05/0.

\end{acknowledgements}

\clearpage


\begin{thebibliography}{}

\bibitem[Costa et al. (1997)]{costa97} Costa, E., Frontera, F., Heise, J., et al, 1997, Nature, 387, 783
\bibitem[Covino et al.(2006)]{Covino2006} Covino, S., Malesani, D., Tagliaferri, G., et al. 2006 Il Nuovo Cimento C, 121, 1171
\bibitem[Klotz et al.(2005)]{Klotz2005} Klotz, A., Bo\"er, M., Atteia, J.L., et al., 2005, A\&A 439, L35
\bibitem[Klotz et al.(2008)]{klo08} Klotz, A., Gendre, B., Stratta, G., et al., 2008, A\&A, 483, 847
\bibitem[Zhang et al. (2003)]{zhang03} Zhang, B., Kobayashi, S., Meszaros, P., 2003, ApJ, 595, 950
\end{thebibliography}
\end{document}